\def\aap{A\&A}

\def\apj{ApJ}
\def\apjl{ApJL}
\def\apss{ApSS}

\def\mnras{MNRAS}

\def\deg{$^\circ$}
\documentclass[letter]{aa}  
\usepackage{graphicx}
\usepackage{txfonts}
\begin{document}

\title{A dense disk of dust around the born-again Sakurai's object}

\authorrunning{Chesneau et al.}
\titlerunning{A dense disk of dust around the Sakurai's object}

   \author{O. Chesneau
          \inst{1}
          \and
          G.C. Clayton
          \inst{2}
          \and F. Lykou
           \inst{3}
          \and
          O. De Marco\inst{4}
          \and
          C.A. Hummel
          \inst{5}
          \and\\
          F. Kerber\inst{5}
          \and
          E. Lagadec\inst{3}
          \and
          J. Nordhaus\inst{6}
          \and
          A.A. Zijlstra\inst{3}
           \and
          A. Evans\inst{7}
    				\fnmsep \thanks{Based on observations made with the Very Large Telescope Interferometer at Paranal Observatory under program 079.D-0415}
          }

   \offprints{O. Chesneau}

   \institute{Observatoire de la C\^{o}te d'Azur-CNRS-UMR 6203, Dept Gemini,
Avenue Copernic, F-06130 Grasse, France\\
              \email{olivier.chesneau@ob-azur.fr}%
              \and
Department of Physics \& Astronomy, Louisiana State University, Baton Rouge, LA 70803, USA
              \and
Jodrell Bank Centre for Astrophysics, The A. Turing Building, The Univ. of Manchester, Oxford Rd, Manchester, M13 9PL, U.K.
\and
Department of Astrophysics, American Museum of Natural History, Central Park West at 79th Street, New York, NY 10024, USA
\and
European Southern Observatory, Karl-Schwarzschild-Strasse 2 D-85748 Garching bei M{\"u}nchen, Germany
\and
Department of Astrophysical Sciences, Princeton University, Princeton, NJ 08544 USA
\and
Astrophysics Group, Lennard-Jones Laboratories, Keele University, Staffordshire ST5 5BG, UK}
   \date{Received ;accepted }

% \abstract{}{}{}{}{} 
% 5 {} token are mandatory
 
  \abstract
   {In 1996, Sakurai's object (V4334 Sgr) suddenly brightened in the centre of a faint Planetary Nebula (PN). This very rare event was interpreted as the reignition of a hot white dwarf that caused a rapid evolution back to the cool giant phase. From 1998 on, a copious amount of dust has formed continuously, screening out the star which has remained embedded in this expanding high optical depth envelope. }
   {The new observations, reported here, are used to study the morphology of the circumstellar dust in order to investigate the hypothesis that Sakurai's Object is surrounded by a thick spherical envelope of dust.}
   {We have obtained unprecedented, high-angular resolution spectro-interferometric observations, taken with the mid-IR interferometer
MIDI/VLTI, which resolve the dust envelope of Sakurai's object.}
   {We report the discovery of a unexpectedly compact (30 x 40 milliarcsec, 105 x 140 AU assuming a distance of 3.5 kpc), highly inclined, dust disk. We used Monte Carlo radiative-transfer
simulations of a stratified disk to constrain its geometric and physical parameters, although such a model is only a rough approximation of the rapidly evolving dust structure. Even though the fits are not fully satisfactory, some useful and robust constraints can be inferred. The disk inclination is estimated to be 75\deg$\pm$3\deg with a large scale height of 47$\pm$7\,AU. The dust mass of the disk is estimated to be $6 \times 10^{-5}M_{\odot}$.  The major axis of the disk (132\deg$\pm$3\deg) is aligned with an asymmetry seen in the old PN that was re-investigated as part of this study. This implies that the mechanism responsible for shaping the dust envelope surrounding Sakurai's object was already at work when the old PN formed.%, some 10000 years ago.
}
   {}

                
%\titlerunning{The dusty discs in the core of the PNe Mz 3 and M2_9}
   \maketitle
%
%________________________________________________________________

\section{Introduction}
Sakurai's Object (V4334 Sgr), first detected in 1996, is the central star of a planetary nebula (CSPN), that experienced a Very Late Thermal Pulse (VLTP), a helium-shell flash while on the white dwarf cooling track that can influence the late evolution of low-mass stars (\cite{2001ApJ...554L..71H,2003ApJ...583..913L,2005Sci...308..231H}). The extended very faint planetary nebula (PN) bears witness to the previous evolution of this star, confirming that the latest large mass ejection during the PN phase occurred several thousands years ago (\cite{1999A&A...344L..79K}). This new 'final flash', which began 12 years ago, returned the star very briefly to the Asymptotic Giant Branch (AGB) stage, explaining why these sources are often called 'Born-again' objects. As such, a final flash is astronomically very brief (only a few tens of years), so that the observation of such an event is rare. Only the event experienced by V605\,Aql can be considered as directly comparable (\cite{2006ApJ...646L..69C,1997AJ....114.2679C}). V605\,Aql underwent a final flash in 1917 and appears today to be still embedded in a disk-like dust structure (\cite{2006ApJ...646L..69C,2008A&A...479..817H}).

The analysis of the very rapid evolution of the spectral energy distribution (SED) of Sakurai's Object (Eyres et al. 1999; Kerber et al. 1999; Tyne et 2002; K{\"a}ufl et al. 2003; Eyres et al. 2004; Evans et al. 2006; Worters et al. 2008) has suggested that the average dust grain size has been increasing and probably the rate of dust formation has been increasing as well. The results were mainly based on modeling of radiative transfer in a spherical dust shell. Tyne et al. (2002) suggested that the source was expanding and might be resolvable by 8-10m class telescopes in the mid-IR. Kerber et al. (2002) discovered of a fast outflow from the source and suggested a bipolar morphology for the fast moving gas seen in optical spectra. Recently, it has been suspected that there is a strong asymmetry in the circumstellar shell around Sakurai's Object, but the dust shell itself had yet to be resolved in the optical or near-IR (\cite{2006MNRAS.373L..75E,2007A&A...471L...9V}). 

Observations using the Very Large Telescope Interferometer are presented in Section 2. In Section\,3 we derive some physical parameters of the dust shell using 2D radiative-transfer models and we discuss the results in Section\,4.

%__________________________________________________________________

\section{Observations}
The source was observed in June 2007 with MIDI (Leinert et al. 2003, Ratzka et al. 2003) the mid-IR recombiner of the Very Large Telescope (VLT). The MIDI/VLT Interferometer operates like a classical Michelson interferometer combining the mid-IR light (N band, 7.5-13.5 $\mu$m) from two VLT Unit Telescopes (8.2m, UTs). A typical MIDI observing sequence was followed, as described in Ratzka et al. (2007). MIDI provided single-dish acquisition images with a spatial resolution of about 250 mas at 8.7 $\mu$m, a flux-calibrated spectrum at low spectral resolution (R=30) and 6 visibility spectra from the source. The observations were performed in High\_Sens mode, implying that the photometry of the source is recorded subsequently to the fringes. The visibility errors range from 8\% to 15\%. The accuracy of the absolute flux calibration is better than 10\%. The log of the observations is given in Table~\ref{table-log}. We used two different MIDI data reduction packages: MIA,
developed at the Max-Planck-Institut f\"ur Astronomie, and EWS,
developed at the Leiden Observatory (MIA+EWS\footnote{Available at http://www.strw.leidenuniv.nl/$\sim$nevec/MIDI/index.html}, ver.1.5.1).

 \begin{table}
\begin{caption}
{Observing log
}\label{table-log}
\end{caption}
\begin{tabular}{llccc}\hline\hline 
OB   & Time  & Base & \multicolumn{2}{c}{Projected baseline}  \\
 & & & Length  & PA   \\
 &&&[metre] & [degrees] \\
\hline
Sak-1 & 2007-06-29T01 & U2 - U3 & 41.9 & 14.6  \\
Sak-2  & 2007-06-29T02 & U2 - U3 & 45.9 & 37.1 \\
Sak-3  & 2007-06-29T06 & U2 - U3 & 44.4 & 48.4  \\
Sak-4  & 2007-06-30T01 & U3 - U4 & 51.3 & 101.0  \\
Sak-5  & 2007-06-30T06 & U3 - U4 & 54.7 & 124.3  \\
Sak-6  & 2007-06-30T08 & U3 - U4 & 39.2 & 159.5  \\
\hline
\end{tabular}
{\tiny Calibrators: HD152334 K4III 3.99$\pm$0.07\,mas, HD163376 M0III 3.79$\pm$0.12\,mas, HD169916 K1III 3.75$\pm$0.04\,mas, HD177716 K1III 3.72$\pm$0.07\,mas.}
\end{table}

\begin{figure*}
 \centering
\includegraphics[height=7.5cm]{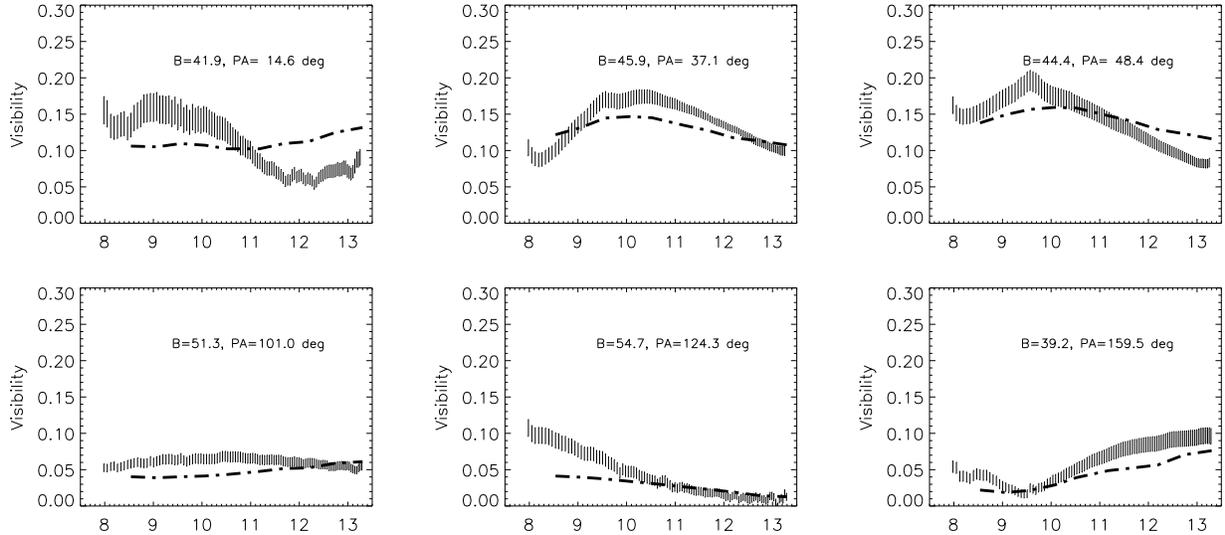}
\hfill
 \caption[]{MIDI visibilities with their error bars, compared with the best model (dashed-dotted lines, see Table\,~\ref{tab:modparam}). The $\chi^2$ values of the fits of 16.5, 7.1, 7.8, 7.6, 5.2, 4.6. The reduced $\chi^2$ for the full data set is 8.2.  \label{fig:MIDIvis}}
\end{figure*}

The MIDI images show that the core is unresolved at 8.7 $\mu$m, implying that its angular diameter must be less than 150-200 mas. The mean level of the visibility spectra is low, indicating that the source is well resolved by MIDI/VLTI  (typical scale 30 x 40 mas), but the spectral modulation indicates a complex source (see Fig.\ref{fig:MIDIvis}). Significant differential phases ($\pm$10-20$^\circ$ peak-to-peak) were observed for all baselines except Sak.4 and Sak.5 (Table~\ref{table-log}), which exhibit phases close to zero ($\pm$5$^\circ$).

\begin{figure}
 \centering
\includegraphics[width=8.2cm]{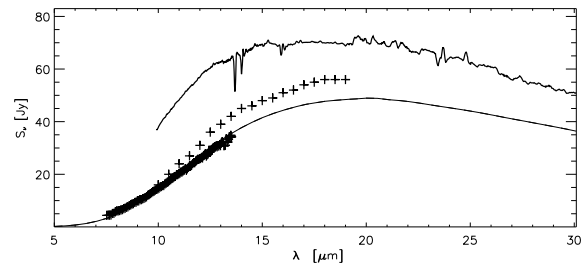}
 \caption[]{MIDI spectrum (June 2007; thick crosses) compared with the Spitzer spectra  obtained in April 2005 (upper solid curve; Evans et al. 2006) and May 2007 (crosses; Evans et al. 2009, in preparation). The smooth solid line is the flux from the best model assuming a distance of 3.1 kpc.
\label{fig:SED}}
\end{figure}

\begin{figure}
 \centering
\includegraphics[width=7cm]{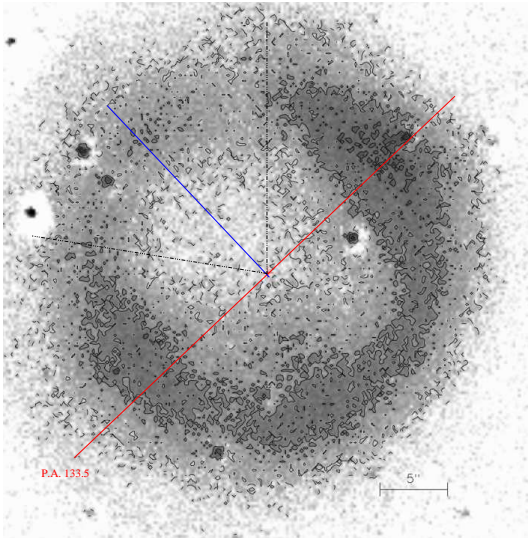}
\hfill
 \caption[]{Continuum subtracted [OIII] contour plot of the old PN recorded in 2002. The region oriented at P.A.=130$^\circ$ is brighter and a hole is visible in the North-East part of the roughly spherical nebula, with an half-aperture angle of $\sim$45$^\circ$ . North is up and East is left. \label{fig:PN}}
\end{figure}

The data are complemented by two Spitzer/IRS spectra obtained in April 2005 and May 2007 (\cite{2006MNRAS.373L..75E}, Evans et al. 2009, in preparation), and a continuum-subtracted [O III] image of the surrounding PN acquired in October 2002 with the instrument VLT/FORS1 (\cite{2005Sci...308..231H}). Fig.\ref{fig:SED} shows the Spitzer and MIDI spectra which are characterized by a steeply rising flux toward the long wavelengths. The [OIII] image of the PN is shown in Fig~\ref{fig:PN}. The appearance of the nebula is almost  circular, with weak but significant spatial structure. 

\section{Physical parameters of the disk}
 The large, spectrally dependent variations of the MIDI visibilities could be interpreted in the context of double point or ring structures. Simple models fail to account for the complex visibilities observed, although the difference between the mean level of the visibilities from the baseline UT2-UT3, compared to those obtained with UT3-UT4 suggest a flattened structure with an orientation of the major axis of approximately 110-140$^\circ$. The differential phases point to a significant asymmetry in the direction of the minor axis, i.e. $\sim$45$^\circ$.

\begin{table}
	\caption{Model parameters. }
	\label{tab:modparam}
	\centering
		\begin{tabular}{ccc}
		\hline
		 & Parameters & Parameter range\\
		\hline
		$T_{\rm eff}$ (K) & $12000$  & -\\
		Luminosity $(L_{\odot})$ & $10000$& -\\
		Distance (kpc)& $3.5$&  -\\
		\hline
		Inclination ($^{\circ}$) & $75$  & 3\\
		PA angle ($^{\circ}$)& 132  &3\\
		Inner radius (AU) &  65 & 10\\
		Outer radius (AU) &  500& No constraint \\
		$\alpha$ & 2.0  & 0.1\\
		$\beta$ & 1.0 & 0.1\\
		$h_{100AU}$ (AU)&47 &7\\
%		Density (g.cm$^{-3}$)&  2.7 & - \\
		Dust mass $(M_{\odot})$&  $6\times 10^{-5}$ & $3\times 10^{-5}$\\
 		\hline
		 
		\end{tabular}
\end{table}
We made use of the continuum Monte Carlo radiative transfer code MC3D (\cite{Wolf2003}; see also \cite{Wolf1999}), that solves the radiative transfer problem self-consistently. We use a classical model of a stratified disk of dust (Chesneau et al. 2006, 2007 and references therein). The dust density follows a 2D law (both radial and vertical):
\begin{equation}		\rho(r,z) = \rho_{0}\left(\frac{R_{\star}}{r}\right)^{\alpha} \exp\left[-\frac{1}{2}\left(\frac{z}{h(r)}\right)^{2}\right]
\end{equation}	
where $r$ is the radial distance in the midplane of the disk, $\alpha$ is the density parameter in the midplane,
	$R_{\star}$ is the stellar radius and the disk scale height, $h(r)$, is given by $h(r)=h_{0}\left(r/R_{\star}\right)^{\beta}$, where $h_{0}$ is the scale height of reference, and $\beta$ is the vertical density parameter.

The absence of spectral features in the IR indicates that the dust is dominated by amorphous carbon grains. We assume the standard interstellar grain size distribution (Mathis et al. 1977) with spherical grain radii extending from $0.005$ to $1~\mu$m. We estimate the outer radius of the disk to be 500\,AU consistent with an expansion since 1997 at a maximum velocity of 100~km~s$^{-1}$. The initial parameters ($T_{\rm eff}$ of the source and dust mass) were chosen following van Hoof et al. (2007). The outputs from the code are described in Chesneau et al. (2006, 2007). 
	The distance, luminosity, and temperature of the central star are uncertain by a large factor (Jacoby et al. 1998). We first fixed the distance to be 1.5 kpc, but no satisfactory models could be found. By increasing the distance, the fit of the visibility curves and the SED steadily improved and we achieved the best fit model for a distance of 3.5 kpc, a value close to the distance favored by van Hoof et al. (2007). We note, however, that we did not explore the parameter space for larger distances and that the distance chosen is by no means a strong constraint. Our priority in this study was to obtain the best possible definition of the the geometry of the source, by fitting the visibility curves. We were not able to find a model with a good match to both the visibility curves and the SED. The best model, presented here, is the one showing the highest flux although it is still too low. Some models with slight modifications of the parameters provided better fits to the visibilities (best $\chi^2 \sim$6). Given the very short timescale to form the disk, and the large scale height inferred, a stratified disk might not be the best model to account for the density profile of the dust. A slowly expanding torus (\cite{Peretto2007}) should be investigated in the future.

\section{Results and discussion}
The MIDI observations provide direct evidence for a thick, highly-inclined disk or torus (which therefore efficiently screens the central source), whose parameters are described in Table 2. The orientation of the major axis of the disk (134$\pm$5$^\circ$) coincides with the asymmetry detected in the old PN (130$\pm$8$^\circ$). A similar orientation is also mentioned in Kerber et al. (2002). Such an alignment between the disk and PN may not be fortuitous. In the case of V605 Aql, the disk and the planetary nebula A58 share the same major axis (\cite{2008A&A...479..817H}). Of importance too, is the low level of asymmetry detected in the old PN compared to what is currently observed in the disk. Is the new mass loss more asymmetric or is the smaller asymmetry of the PN remnant due to evolution and old age?
Another related object worth mentioning is A\,30, a PN that suffered its own VLTP a few thousand years ago (\cite{1995ApJ...449L.143B}). The inner nebula exhibits a thin equatorial disk, currently being eroded by the fast wind of the central star, whereas its old PN is perfectly round. \par

\begin{figure*}
 \centering
\includegraphics[height=6cm]{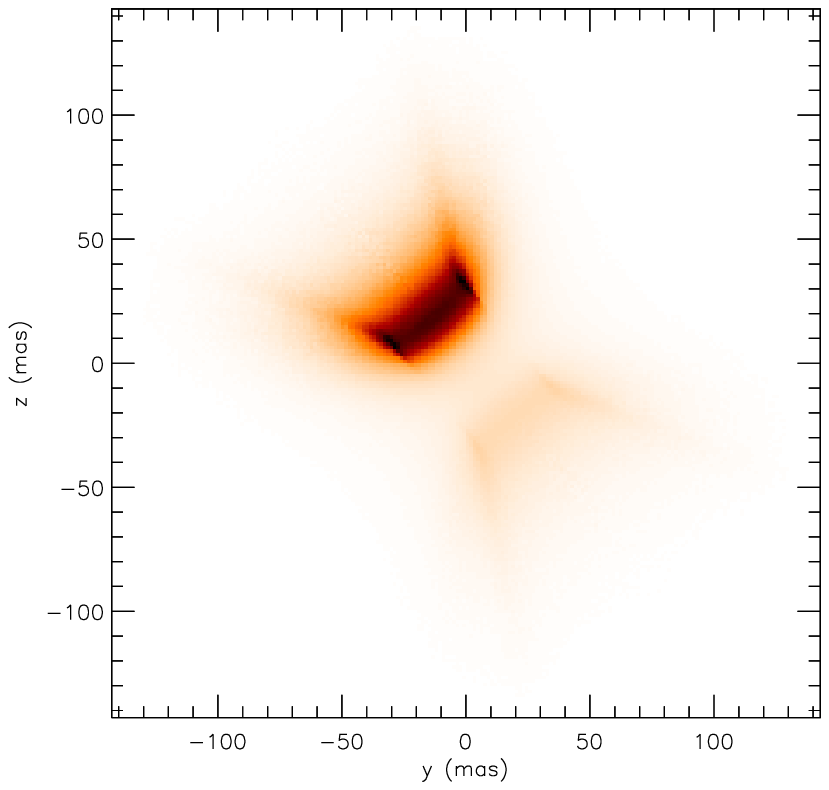}
\includegraphics[height=6cm]{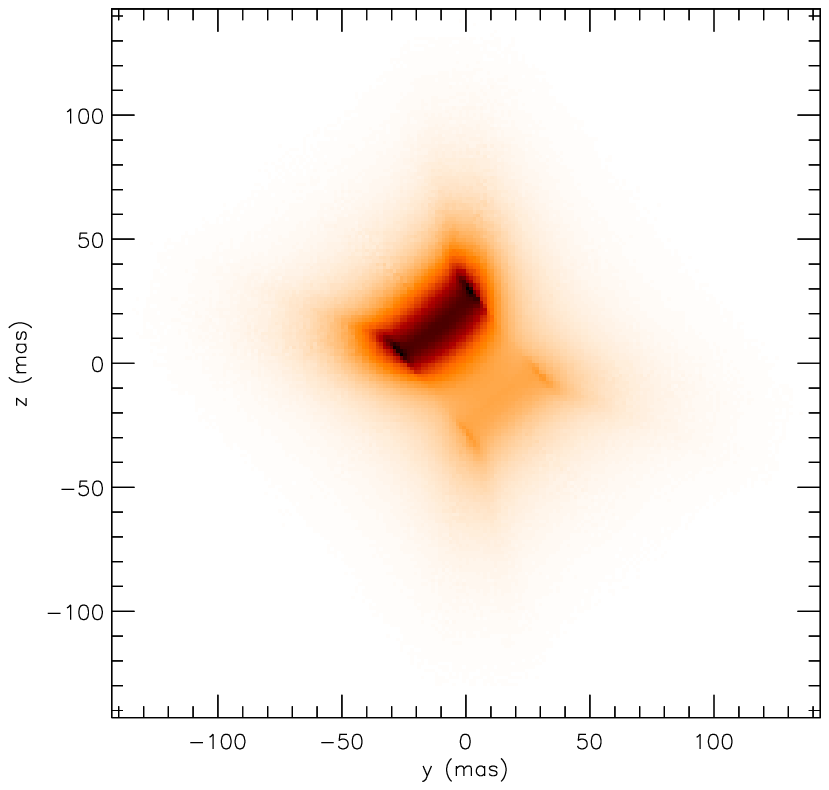}
\includegraphics[height=6cm]{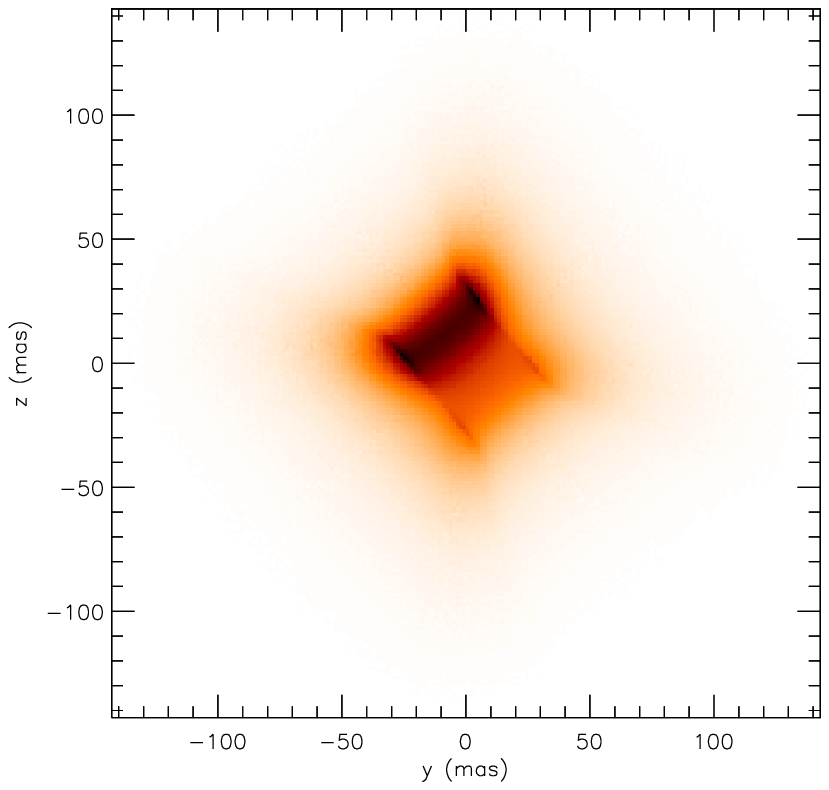}
\hfill
 \caption[]{Flux distribution of the model at 8, 10.5 and 13 $\mu$m. A small amount of light emerges at 8 $\mu$m while the flux comes from deeper in the disk at 13 $\mu$m. 
\label{fig:maps}}
\end{figure*}

The disk is very optically thick and at 8 $\mu$m, most of the light emerges from a small region above the North pole (see Fig.\ref{fig:maps}). The scale-height of the disk/torus is very large, limiting the opening angle of the polar regions, reminiscent of the long-lived `wall' structure seen in the binary post-AGB HR 4049 \cite{Dominik03}.  Spherical models of the dust shell explain the shift of the peak of the SED toward longer wavelengths by suggesting that the inner boundary has receded from the star (\cite{2003A&A...406..981K,2006MNRAS.373L..75E}). If this indicated a cessation of mass loss and dust formation, then the optical depth of dust would drop very quickly, assuming that the dust is moving radially away from the star. 
The estimated size of the Sakurai's Object disk from our observations in 2007 is small compared with previous estimates, and implies that the high velocities inferred from spectral lines may not be consistent with the dust expansion.
If the expansion velocity were consistent with the rapid motions observed for CO  and optical lines, the disk should already be resolvable by 8-m class telescope in the infrared (\cite{2004MNRAS.350L...9E, 2002ApJ...581L..39K, 2008arXiv0810.4556W}). \par

The fast spectral evolution of the source toward lower temperatures, as witnessed by the Spitzer and MIDI spectra, must be interpreted by a growing opacity of the disk, whether by accumulation of new material or by grain growth. Only the continuous production of new dust and/or grain growth can explain the shift of the SED of the disk toward longer wavelengths. A more rapid expansion of the disk as an explanation for the SED evolution has been ruled out by the smaller disk size reported here. Since the dust on the inner rim of the disk is close to the sublimation temperature, and the outer radius is not well constrained by the N-band measurements, a dust expansion rate cannot be accurately estimated. Keplerian motion (i.e. with a very low radial expansion) of the dust in the disk cannot be excluded.\par
%The fact that the visual (V, R, I) magnitudes have reached a plateau close to 20-25 magnitude is a direct consequence %of the disk-like geometry. The visible flux from the star is most probably reflected light. 

The discovery of a flattened structure such as a disk is of great importance to the future interpretation of the outburst of Sakurai's Object and the VLTP phenomenon, as much of the previous work is based on the assumption of a spherical ejection. Disk models may do a better job of fitting the observed parameters. A well-known consequence of applying a 1D radiative transfer code in the dust to a 2D astronomical source is that the dust is put much farther than in reality, biasing strongly all the inferred parameters.
%Many of the features described in the literature can be explained if they emerge from various co-latitudes with respect to the disk axis. 


The formation of a disk/torus could be explained by two classes of models: those involving an intrinsic asymmetry of the ejection, and those involving a binary companion. The models must also account for the asymmetry seen in the surrounding PN which implies that the same asymmetry has been acting on the mass loss for 10\,000~yr. 
Binarity can naturally explain the disk-like geometry and also the fast wind features (such as the He I $\lambda$10830 line and the hole at P.A.$\sim$45$^\circ$). The companion would likely be a very low-mass object which has survived the previous evolution of the system (\cite{2007ApJS..171..206M}). The fact that the asymmetry of the disk is much more pronounced in the recent VLTP event could be a consequence of the smaller mass of the ejecta. It is important to stress that in this scenario the original ejection might have been {\it spherical}, but with the asymmetry being enhanced by the influence of the companion. The short timescale for the disk formation is a severe constraint and may imply that the orbital period of the companion is short (i.e. typically a few months). 
Given their similarities, any solution to the asymmetry of the Sakurai's Object dust envelope will also apply to its twin, V605\,Aql (\cite{2006ApJ...646L..69C}). In both cases, we seem to have a thick disk of dust
preventing the star from being seen directly. In both cases too, some structure in the surrounding PNe are aligned with the axis of the disk. 
 
The small rate of expansion of the disk is a strong argument for a binary interaction. A VLTP happens in 10-20\% of post-AGB stars. A companion, originally far from the primary ($\sim$5-10\,AU) so to avoid mass transfer and a common envelope phase could be responsible for the outflow shaping. During the initial AGB phase, the spherical wind from the primary star interacts with the companion and is focused toward the equatorial plane.  Initially the asymmetry is strong. However, after 10$^4$ years, the asymmetry slowly disappears as the nebula expands into the ISM. When the VLTP occurs, the resulting mass-loss also interacts with the companion and is once again focused towards the equatorial plane. Given the mass ejected, the companion might also migrate inward and may provide enough angular momentum to stabilise the ejecta in the form of a torus (\cite{2006MNRAS.370.2004N,Peretto2007, edgar2008}). Given the disk inclination and high optical depth, the detection of such a companion will be a difficult challenge for the foreseeable future.
%A VLTP happens in 10-20\% of the times to post-AGBs, a significant fraction of which might be in binary systems. 

%\begin{acknowledgements}
%\end{acknowledgements}

\end{document}